\PassOptionsToPackage{table}{xcolor}
\documentclass[runningheads]{llncs}
\usepackage[T1]{fontenc}
\usepackage{graphicx}
\usepackage{booktabs}

\usepackage{amsmath,amsxtra,amssymb,latexsym, amscd}
\usepackage{times}
\usepackage{latexsym}
\usepackage{graphicx}
\usepackage{multirow}
\usepackage{pgf-pie}
\usepackage{tablefootnote}
\usepackage{tabularx}
\usepackage{adjustbox}
\usepackage{float}
\usepackage{graphicx}
\usepackage{caption}
\usepackage{lipsum}
\usepackage{orcidlink}
\usepackage{bbm}
\usepackage{flushend}
\usepackage{longtable}
\usepackage{makecell}
\usepackage{array,xcolor}
\usepackage{soul,soulutf8}
\usepackage{subcaption}
\usepackage{pgfplots}
\pgfplotsset{compat=1.18}
\usetikzlibrary{patterns}
\usepackage{booktabs}
\usepackage{siunitx}
\usepackage[table]{xcolor} 

\usepackage[mathscr]{eucal}

\usepackage[T5]{fontenc}
\usepackage{CJKutf8}
\usepackage{hyperref}

\usepackage[style=numeric-comp, sorting=none, maxbibnames=99]{biblatex}
\begin{document}

\title{Optimizing Legal Document Retrieval in Vietnamese with Semi-Hard Negative Mining}

\titlerunning{Optimizing Legal Doc. Retrieval in Vietnamese w/ Semi-Hard Negative Mining}

\author{
Van-Hoang Le\orcidlink{0009-0000-1924-7060}\inst{1,4} \and
Duc-Vu Nguyen\orcidlink{0000-0003-0072-2524}\inst{3,4}\thanks{Corresponding author} \and \\
Kiet Van Nguyen\orcidlink{0000-0002-8456-2742}\inst{2,4} \and
Ngan Luu-Thuy Nguyen\orcidlink{0000-0003-3931-849X}\inst{2,4}
}

\authorrunning{Van-Hoang Le et al.}

\institute{Faculty of Computer Science, University of Information Technology, Ho Chi Minh City, Vietnam \and Faculty of Information Science and Engineering, University of Information Technology, Ho Chi Minh City, Vietnam \and Laboratory of Multimedia Communications, University of Information Technology, Ho Chi Minh City, Vietnam \and
Vietnam National University, Ho Chi Minh City, Vietnam
\email{\{22520465\}@gm.uit.edu.vn~\{vund,kietnv,ngannlt\}@uit.edu.vn}}

\maketitle

\begin{abstract}
Large Language Models (LLMs) face significant challenges in specialized domains like law, where precision and domain-specific knowledge are critical. This paper presents a streamlined two-stage framework consisting of Retrieval and Re-ranking to enhance legal document retrieval efficiency and accuracy. Our approach employs a fine-tuned Bi-Encoder for rapid candidate retrieval, followed by a Cross-Encoder for precise re-ranking, both optimized through strategic negative example mining. Key innovations include the introduction of the Exist@m metric to evaluate retrieval effectiveness and the use of semi-hard negatives to mitigate training bias, which significantly improved re-ranking performance. Evaluated on the SoICT Hackathon 2024 for Legal Document Retrieval, our team, 4Huiter, achieved a top-three position. While top-performing teams employed ensemble models and iterative self-training on large bge-m3 architectures, our lightweight, single-pass approach offered a competitive alternative with far fewer parameters. The framework demonstrates that optimized data processing, tailored loss functions, and balanced negative sampling are pivotal for building robust retrieval-augmented systems in legal contexts.
\end{abstract}

\section{Introduction}
Recent advancements in LLMs, such as ChatGPT, have revolutionized general-purpose question answering. However, applying these models to specialized domains like law remains challenging due to the complexity of legal language, which requires precise interpretation and contextual awareness. Retrieval-Augmented Generation (RAG) \cite{RAG} has emerged as a promising solution by combining retrieval systems with LLMs to improve accuracy. The effectiveness of RAG depends heavily on the retrieval component, especially in legal contexts where document length, specificity, and semantic nuance make information extraction more difficult.

This study addresses these challenges through the context of the SoICT Hackathon 2024 Legal Document Retrieval competition, which tasked participants with developing efficient retrievers for a corpus of 261,446 legal texts. While other competitors relied on ensembling, we propose a two-stage pipeline consisting of a Bi-Encoder for fast retrieval and a Cross-Encoder for refined re-ranking. Our framework introduces two critical advancements: (1) the Exist@m metric to ensure candidate relevance during retrieval, and (2) semi-hard negative mining to balance training data and improve re-ranking robustness.

Experiments demonstrate that a fine-tuned Vietnamese Bi-Encoder outperforms lexical methods like BM25, achieving 97\% Exist@90. For re-ranking, semi-hard negatives mined from Bi-Encoder candidates resulted in a 23\% relative improvement over baseline models, with a 79.11\% MRR@10 score on the evaluation dataset and a 77.54\% MRR@10 score on the private test. These results highlight the importance of domain-adaptive training, strategic negative sampling, and pipeline optimization. Our contributions offer actionable insights for legal information retrieval systems and emphasize the potential of tailored retrieval architectures in addressing the limitations of LLMs for specialized domains.

\section{Related Work}

Information Retrieval (IR) systems have evolved along two main lines: lexical and semantic approaches. Lexical methods, such as TF-IDF \cite{TF-IDF} and BM25 \cite{BM25}, rely on exact word matching between queries and documents. While they are fast and simple, they struggle with linguistic variation, such as synonyms, polysemy, or contextual meaning shifts, making them less effective for deeper semantic understanding.

To address these limitations, neural network–based semantic models have been introduced. Methods like Sentence-BERT (SBERT) \cite{reimers2019sentencebert} and Universal Sentence Encoder (USE) \cite{USE} encode text into high-dimensional embeddings that capture sentence meaning. These models handle paraphrasing and domain-specific language more effectively, but their higher computational cost often necessitates hybrid systems, especially in large-scale applications.

Legal document retrieval presents additional challenges due to the length, complexity, and formal structure of legal texts. Early work by \citeauthor{first} \cite{first} explored document embeddings and deep learning to improve retrieval. Later, \citeauthor{second} \cite{second} introduced attention mechanisms to highlight key text segments. More recent advances include multi-stage architectures, such as the system proposed by \citeauthor{third} \cite{third}, which combines BM25 retrieval, BERT-based re-ranking, and LLM prompting for final scoring.

Our work follows the multi-stage design but focuses on efficiency and robustness. Rather than using ensemble models, which are often resource-intensive, we develop a lightweight two-stage pipeline: a Bi-Encoder for retrieval and a Cross-Encoder for re-ranking. Both stages are fine-tuned with legal data and enhanced through domain adaptation and semi-hard negative sampling. This strategy improves generalization while maintaining scalability, making it well-suited for real-world legal IR systems.

\section{Background}

\subsection{Retrieval Using Bi-Encoder Models}

A Bi-Encoder employs a dual-tower architecture where the query and document are encoded independently into vector embeddings, typically using models like Sentence-BERT \cite{reimers2019sentencebert}. These embeddings can be rapidly compared via cosine similarity, making Bi-Encoders efficient for retrieving top candidates (e.g., the top 90) from large corpora. However, because the encoding is independent, the model may overlook fine-grained interactions between the query and document. In this work, we use a Vietnamese Bi-Encoder\footnote{Pretrained Vietnamese legal-domain Bi-Encoder model available at: \url{https://huggingface.co/bkai-foundation-models/vietnamese-bi-encoder}} trained on legal texts to improve domain-specific retrieval.

\subsection{Re-ranking with Cross-Encoder Models}

A Cross-Encoder processes the query and document together using a single-tower architecture. This allows the model to capture richer semantic interactions, leading to more accurate relevance scores \cite{reimers2019sentencebert}. However, since it evaluates each query-document pair individually, it is computationally expensive and typically used to re-rank a small number of candidates selected by a faster retriever.

\subsection{Role of Negative Examples in Training}

Training an effective retriever requires both relevant and non-relevant (negative) documents. Negative examples help the model learn to differentiate between correct and incorrect matches. If the negative samples are too easy, the model gains little from the contrast. On the other hand, if they are too difficult, they may confuse the model. Semi-hard negatives, which are similar to the correct answers but still incorrect, strike a good balance and have been shown to improve performance \cite{gillick2019learning}.

\subsection{Contrastive Loss with Multiple Negatives}

Multiple Negatives Ranking Loss is a contrastive objective that increases similarity between matching pairs while treating all other in-batch samples as negatives \cite{multiplenegativerankingloss}. It reduces bias in datasets dominated by positive pairs and is especially effective in legal question answering, where distinguishing correct from nearly correct answers is crucial.

\section{Exploratory Data Analysis}

The dataset used in this study originates from the SoICT Hackathon 2024 competition and consists entirely of legal texts written in Vietnamese. It is divided into three main files: \texttt{train.csv}, \texttt{corpus.csv}, and \texttt{public\_test.csv}. In this section, we describe the structure and key characteristics of each file to provide context for subsequent model development and evaluation.

\subsection{Structure and Characteristics of the Training Data (\texttt{train.csv})}


\begin{figure}[H]
  \centering
  \begin{subfigure}{0.45\linewidth}
    \centering
    \begin{tikzpicture}
      \begin{axis}[
          ybar,
          ymin=0,
          xlabel={Number of tokens},
          ylabel={Questions},
          symbolic x coords={{3--10},{11--25},{26--59}},
          xtick=data,
          nodes near coords,
          nodes near coords align={vertical},
          enlarge x limits=0.25,
          width=1.1\linewidth,
          height=6cm,
        ]
        \addplot+[bar width=18pt]
          coordinates { (3--10,5898) (11--25,89316) (26--59,24242) };
      \end{axis}
    \end{tikzpicture}
    \caption{Most questions ($\approx$75\%) contain 11–25 tokens, while fewer than 5\% are shorter than 10}
    \label{fig:tokens}
  \end{subfigure}
  \hspace{0.5cm}
  \begin{subfigure}{0.45\linewidth}
    \centering
    \begin{tikzpicture}
      \begin{axis}[
          ybar,
          ymin=0,
          xlabel={Document IDs per question},
          symbolic x coords={1,2,3,{4+}},
          xtick=data,
          nodes near coords,
          nodes near coords align={vertical},
          enlarge x limits=0.25,
          width=1.1\linewidth,
          height=6cm,
        ]
        \addplot+[bar width=18pt]
          coordinates { (1,106884) (2,11282) (3,1115) (4+,175) };
      \end{axis}
    \end{tikzpicture}
    \caption{Nearly 90\% of questions reference exactly one document; multi-document questions are rare}
    \label{fig:cids} 
  \end{subfigure}

  \caption{Dataset statistics: token-length profile (left) and document-ID usage per question (right)}
  \label{fig:dataset_stats}
\end{figure}
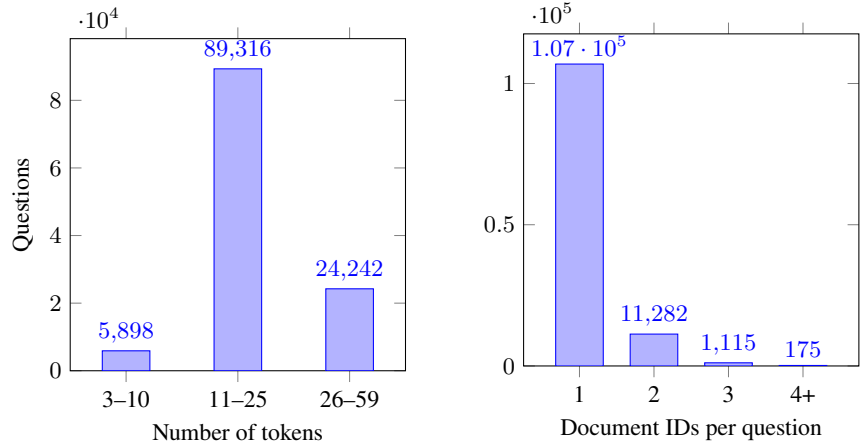

The file \texttt{train.csv} serves as a supervised Question-Answer (QA) dataset intended for model fine-tuning. It contains four columns: \texttt{question}~(query text), \texttt{context}~(answer span), \texttt{cid}~(source-document ID), and \texttt{qid}~(question ID). The \texttt{question} column includes legal questions of varying lengths, ranging from 3 to 59 tokens, with an average of 20 tokens per question. A detailed distribution is shown in Figure~\ref{fig:tokens}.

The \texttt{context} column contains the corresponding answers for each question. Each answer is extracted from the corpus, based on the matching \texttt{cid} values, and is sometimes truncated or slightly modified. Importantly, a single question may have more than one valid answer, so the \texttt{context} field is represented as a list. The \texttt{cid} column refers to the IDs of the documents from which these answers are sourced. As indicated in Figure~\ref{fig:cids}, around 10\,\% of questions are linked to multiple answers. Lastly, the \texttt{qid} column provides a unique identifier for each question in the dataset.

\subsection{Overview of the Legal Document Corpus (\texttt{corpus.csv})}


\begin{figure}[h!]
  \centering
  \begin{tikzpicture}
    \begin{axis}[
        ybar,
        width=\linewidth,        
        height=6.5cm,            
        ymin=0,
        xlabel={Number of tokens},
        ylabel={Documents},
        symbolic x coords={{0--128},{128--256},{256--512},{512--1024},{1024+}},
        xtick=data,
        nodes near coords,
        nodes near coords align={vertical},
        enlarge x limits=0.15,
        bar width=18pt,
      ]
      \addplot+ coordinates {
        (0--128,99786)
        (128--256,81830)
        (256--512,54393)
        (512--1024,23832)
        (1024+,1756)
      };
    \end{axis}
  \end{tikzpicture}

  \caption{Distribution of legal documents by token length}
  \label{fig:doc_token_dist}
\end{figure}
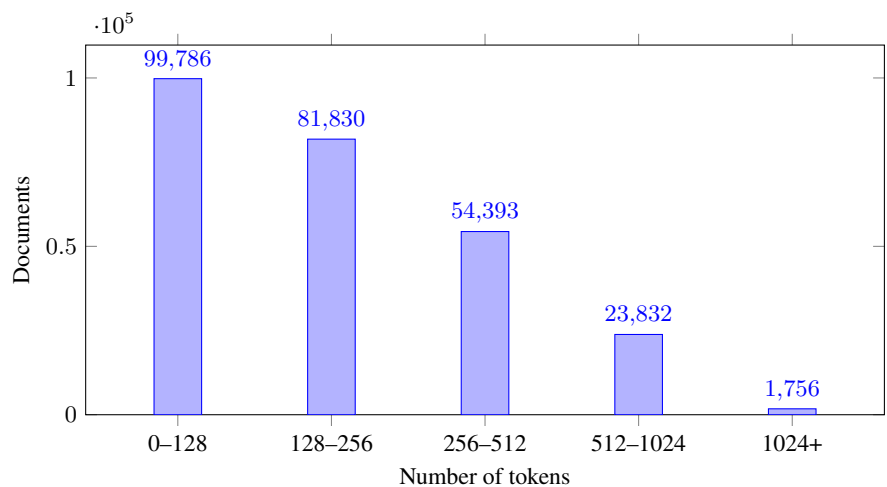

The \texttt{corpus.csv} file contains a large collection of 261,446 legal text segments. It consists of two columns: \texttt{text} and \texttt{cid}. The \texttt{text} column includes the actual legal content, while the \texttt{cid} column assigns a unique identifier to each entry. The documents vary greatly in length. While some segments contain no tokens (and may be considered noise), the longest reaches 50,453 tokens. On average, a document contains about 230 tokens. Figure~\ref{fig:doc_token_dist} presents the token-length distribution.

\subsection{Evaluation Set Overview (\texttt{public\_test.csv})}

The \texttt{public\_test.csv} file contains 10,000 unlabeled questions. These questions range from 4 to 73 tokens in length, with an average of 20 tokens. However, since the file does not include any labels or reference answers, it is not used for training or evaluation in this study.

\subsection{Key Observations and Preprocessing Implications}

An analysis of the data tables above reveals several important considerations for model design. First, both documents and questions are generally fewer than 512 tokens in length. This suggests that embedding models with a maximum sequence length of 512 tokens are well-suited to this task. Nevertheless, it may be beneficial to experiment with models capable of handling up to 1024 tokens, particularly for the longest legal texts.

Second, as shown in Figure~\ref{fig:cids}, a significant portion of questions are linked to multiple answers. Given how BERT-style models operate in QA settings, it is advisable to convert these multi-answer entries into multiple distinct question-answer pairs. This transformation allows the model to better learn answer relevance in a fine-grained manner.

Lastly, although the QA-style format of \texttt{train.csv} is useful for training, it may introduce bias. Models like the Bi-Encoder and Cross-Encoder can overfit to the ``positive'' associations between questions and answers, without learning to reject irrelevant ones. This relevance bias can hinder generalization, especially for unseen queries. Therefore, careful selection of negative samples and thoughtful evaluation strategies are essential for building robust retrieval systems.

\section{Data Preparation and Processing Pipeline}

\begin{figure}[H]
\centering
\includegraphics[width=1\linewidth]{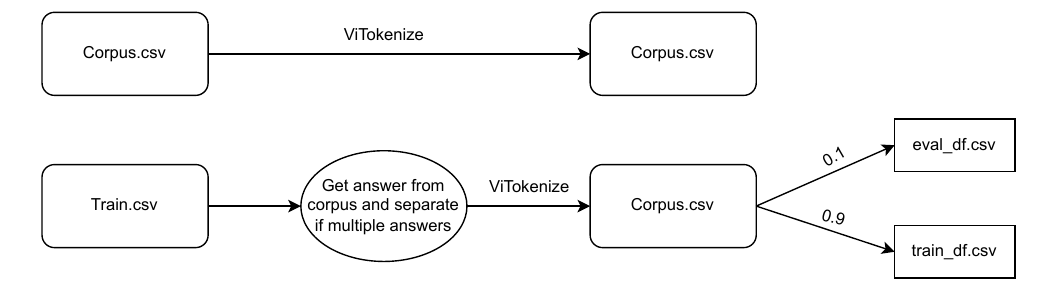}
\caption{Overview of the data processing pipeline}
\label{fig1}
\end{figure}

This section outlines the steps taken to prepare and transform the raw data for model training. The main datasets involved are \texttt{train.csv} and \texttt{corpus.csv}. The overall data processing workflow is illustrated in Figure~\ref{fig1}.

\subsection{Replacing Truncated Answers with Full Legal Documents}

The \texttt{context} column in \texttt{train.csv} often contains shortened or modified excerpts of documents from \texttt{corpus.csv}. To ensure consistency and retain full legal context, we replaced each \texttt{context} entry with the complete document referenced by its corresponding \texttt{cid}. This improves data quality by providing more complete input for training.

\subsection{Separating Multiple Answers per Question}

Some questions have multiple correct answers. To align with standard training formats, where each query maps to one document, we split them into individual question-answer pairs. This results in some questions appearing multiple times, each paired with a different legal text.

\subsection{Vietnamese Word Segmentation with Pyvi}

Vietnamese lacks clear word boundaries, making segmentation a necessary preprocessing step. We applied the Pyvi library\footnote{Pyvi is an open-source Vietnamese tokenizer available at \url{https://github.com/trungtv/pyvi}} to segment both \texttt{train.csv} and \texttt{corpus.csv}. This helps token-based models learn better representations of Vietnamese text, improving retrieval and re-ranking accuracy.

\subsection{Splitting Data for Local Validation}

To enable local model evaluation, we split the processed \texttt{train.csv} into training and validation sets using a 90/10 ratio. This was essential due to submission limits and the lack of labels in \texttt{public\_test.csv}. The validation set allowed us to track performance and tune models effectively during development.

\section{Proposed Methodology: A Two-Stage Retrieval and Re-ranking Framework}

\begin{figure}[H]
\centering
\includegraphics[width=1\linewidth]{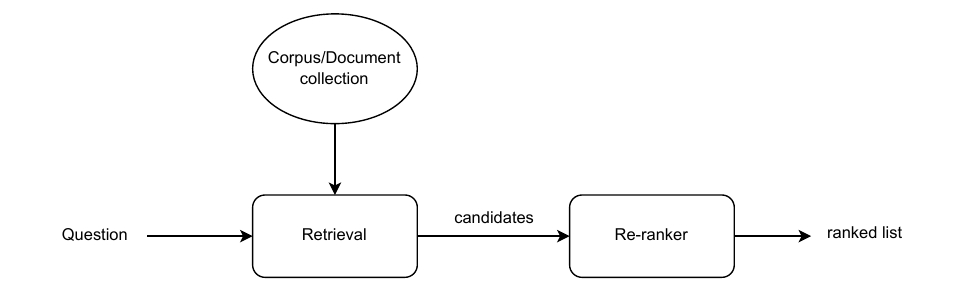}
\caption{Retrieve and re-rank pipeline structure}
\label{fig2}
\end{figure}

In this section, we present our retrieval framework, which follows a well-established two-stage pipeline consisting of an initial retrieval phase using a Bi-Encoder and a re-ranking phase using a Cross-Encoder. Prior work has shown that combining these two components results in effective hybrid retrieval systems for information retrieval (IR) tasks \cite{method1, method2, method3, method4}. A typical structure for such a pipeline is illustrated in the official Sentence-Transformers documentation and reproduced in Figure~\ref{fig2}.

Our implementation follows this general structure, but with careful consideration of model selection, training bias, loss function choice, and evaluation metrics tailored to the legal document retrieval task. Figure~\ref{fig4} summarizes our full fine-tuning and evaluation pipeline.

\subsection{Model Selection and Adaptation through Transfer Learning}

To reduce training time and leverage existing knowledge, we adopt a transfer learning approach. Specifically, we use a pre-trained Sentence-BERT model for retrieval and a pre-trained Cross-Encoder model for re-ranking, both of which have been fine-tuned on Vietnamese language data \cite{bkai, PhoRanker}. These models are further fine-tuned on our legal document dataset to adapt them to the domain-specific vocabulary, structure, and retrieval needs of the task.

\subsection{Training the Bi-Encoder for Initial Candidate Retrieval}
\label{existm}
The Bi-Encoder model is trained using question-answer pairs extracted from the preprocessed \texttt{train.csv} file. We first experimented with \textbf{CosineSimilarityLoss}, but quickly observed a training bias: because the dataset only contains positive examples, the model tended to push all similarity scores toward 1.

To address this, we adopt \textbf{MultipleNegativesRankingLoss}, a contrastive learning approach that treats other in-batch samples as negatives. This encourages the model to distinguish between relevant and irrelevant documents without requiring manually labeled negative examples. Notably, this method becomes more effective with larger batch sizes, as it naturally increases the number of negatives seen by the model.

Although standard retrieval metrics such as MRR@10 could be used to evaluate the Bi-Encoder’s performance, we argue that this is unnecessary when a second re-ranking stage is present. Instead, we propose a simpler and more targeted evaluation metric, called \textbf{Exist@m}, which checks whether at least one correct answer appears among the top-$m$ retrieved candidates:

\begin{equation}
Exist@m = \frac{1}{N} \sum_{i=1}^{N} \mathbbm{1}[\text{Correct answer exists in top } m]
\end{equation}

Here, $N$ is the total number of evaluation queries, and the metric returns 1 for each instance where at least one correct document is found in the top $m$ candidates retrieved by the Bi-Encoder. This metric is particularly useful for pipelines where the re-ranking step will handle final ordering.

\subsection{Fine-tuning the Cross-Encoder for Re-ranking}

After training the Bi-Encoder, we proceed to fine-tune the Cross-Encoder using the same question-answer pairs. However, since the dataset consists only of positive pairs, the Cross-Encoder faces a similar training bias. Unlike the Bi-Encoder, it does not support contrastive loss functions by default, so we must generate high-quality negative samples explicitly.

\begin{figure}[h!]
\centering
\includegraphics[width=1\linewidth]{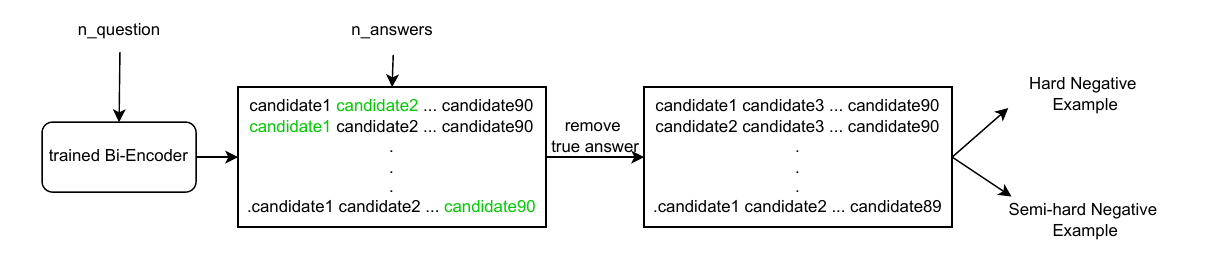}
\caption{Negative example mining using a fine-tuned Bi-Encoder}
\label{fig3}
\end{figure}

\begin{figure}[h!]
\centering
\includegraphics[width=1\linewidth]{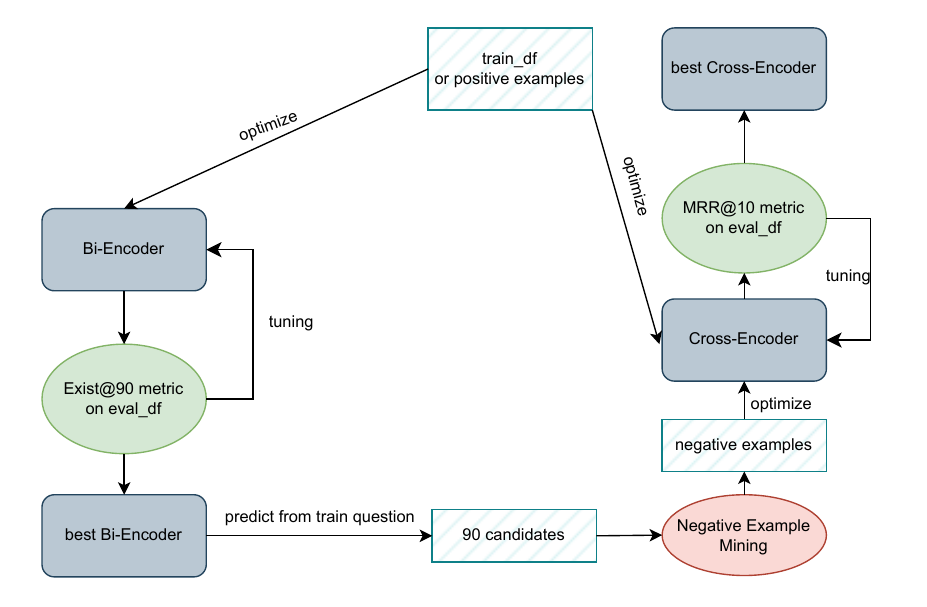}
\caption{End-to-end fine-tuning and evaluation process}
\label{fig4}
\end{figure}

To improve training, we perform \textbf{negative example mining} using the fine-tuned Bi-Encoder. As illustrated in Figure~\ref{fig3}, we explore two types of negative samples:
\begin{itemize}
\item \textbf{Semi-hard negatives}: These are documents that are close to the query in embedding space but are not correct answers. They provide strong training signals and significantly improve re-ranking performance.
\item \textbf{Hard negatives}: These are even closer to the query but tend to introduce noise and often degrade model performance.
\end{itemize}

We generate training data by combining the positive examples from \texttt{train.csv} with the semi-hard negatives mined from the Bi-Encoder. For optimization, we use the \textbf{BCEWithLogitsLoss} (Binary Cross-Entropy Loss with Logits), which is well-suited for binary classification tasks involving relevance scoring.

Model performance is evaluated using \textbf{MRR@10} (Mean Reciprocal Rank at 10), which measures the rank of the first relevant document in the predicted list. A higher MRR@10 score indicates more effective re-ranking.

\subsection{Inference Workflow for Legal Question Answering}

\begin{figure}[h!]
\centering
\includegraphics[width=1\linewidth]{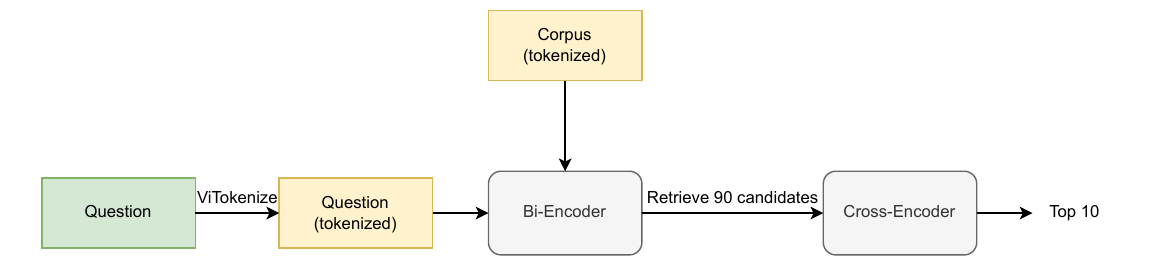}
\caption{Inference process for a legal question}
\label{fig5}
\end{figure}

The final inference pipeline, illustrated in Figure~\ref{fig5}, follows a three-step process. First, the input legal question is tokenized for downstream processing. Next, the tokenized question is passed through the trained Bi-Encoder to retrieve the top 90 candidate documents from the corpus. Finally, each of these candidate documents is evaluated using the Cross-Encoder, which compares them directly to the query. The top 10 documents with the highest relevance scores are returned as the final output.

This hybrid design ensures both speed and accuracy. The Bi-Encoder quickly narrows the search space, and the Cross-Encoder provides precise relevance judgments, making the system suitable for real-time or large-scale legal information retrieval tasks.

\section{Experimental Setup and Evaluation Strategy}

This section presents the experiments conducted to evaluate the effectiveness of our retrieval framework. We focus on both components of the pipeline: the initial retrieval stage using a Bi-Encoder and the re-ranking stage using a Cross-Encoder. We begin by evaluating the retrieval step.

\subsection{Evaluating Candidate Retrieval Using the \textit{Exist@m} Metric}

In our approach, the retrieval component is responsible for selecting 90 candidate documents, which are then re-ranked by the Cross-Encoder. The success of this pipeline depends heavily on whether the correct answer is included in the candidate set. To assess this, we adopt the \textbf{Exist@m} metric (as defined in Subsection~\ref{existm}), which calculates the proportion of cases where at least one correct answer appears among the top-$m$ retrieved candidates.

To examine whether a fine-tuned semantic retriever can outperform a traditional lexical method, we conduct a comparative experiment between the Bi-Encoder and BM25. The Bi-Encoder is fine-tuned using the preprocessed \texttt{train\_df} and \texttt{eval\_df} datasets. We follow the Sentence-Transformers training pipeline and use the \textbf{MultipleNegativesRankingLoss} to optimize performance. The hyperparameters are as follows: a learning rate of $4 \times 10^{-5}$, batch size of 64, no warm-up steps, and a weight decay of 0.02. The model is trained for 7, 9, and 11 epochs to evaluate the impact of training duration.

\begin{table}[h!]
  \centering
  \caption{A quantitative comparison between Bi-Encoder and BM25 models on a Vietnamese legal document dataset. The evaluation framework incorporates two pertinent metrics, Exist@90 and MRR@10, to quantify and contrast the retrieval effectiveness of both models. Results are derived from evaluations conducted on the \texttt{eval\_df} dataset}
  \label{tab:comparison}
  \begin{tabularx}{\textwidth}{Xcc}
    \toprule
    \textbf{Model} & \textbf{Exist@90} & \textbf{MRR@10} \\
    \midrule
    BM25Okapi (default)               & 0.8071 & 0.3437 \\
    BM25Plus (default)                & 0.8250 & 0.3595 \\
    BM25Plus, $k=0.8$, $b=0$          & 0.6902 & 0.1451 \\
    BM25Plus, $k=0.8$, $b=0.75$       & 0.8185 & 0.3582 \\
    BM25Plus, $k=0.8$, $b=1$          & 0.8229 & 0.3660 \\
    BM25Plus, $k=1.2$, $b=0$          & 0.6792 & 0.1307 \\
    BM25Plus, $k=1.2$, $b=0.75$       & 0.8249 & 0.3606 \\
    BM25Plus, $k=1.2$, $b=1$          & 0.8256 & 0.3634 \\
    BM25Plus, $k=2$, $b=0$            & 0.6506 & 0.1090 \\
    BM25Plus, $k=2$, $b=0.75$         & 0.8248 & 0.3559 \\
    BM25Plus, $k=2$, $b=1$            & 0.8215 & 0.3491 \\
    Vietnamese-bi-encoder             & 0.8670 & 0.4607 \\
    Vietnamese-bi-encoder, 7 epochs   & \textbf{0.9748} & \textbf{0.6082} \\
    Vietnamese-bi-encoder, 9 epochs   & \textbf{0.9751} & \textbf{0.6131} \\
    Vietnamese-bi-encoder, 11 epochs  & \textbf{0.9760} & \textbf{0.6222} \\
    \bottomrule
  \end{tabularx}
  \label{tab4}
\end{table}

For comparison, we apply the BM25 method directly to the preprocessed evaluation data. All text is lowercased prior to indexing. We experiment with two configurations: BM25Okapi using default parameters, and BM25Plus with tuned $k_1$ and $b$ values.

This setup enables a fair evaluation of whether a domain-adapted Bi-Encoder can deliver better retrieval performance than strong lexical baselines. Results are analyzed in the following section.

As previously discussed, we adopt \textbf{Exist@90} as the primary metric for selecting the best retrieval model. This metric is particularly suited for pipelines that include a re-ranking stage, as it focuses on whether the correct answer appears within the retrieved candidate set, rather than its exact rank.

Table~\ref{tab4} demonstrates that \textit{Exist@90} is not redundant with traditional metrics like MRR@10; the two do not always correlate, highlighting the importance of task-specific evaluation. In our experiments, the Bi-Encoder consistently outperforms the BM25 baseline. The pre-trained Bi-Encoder model~\cite{bkai}, which has been trained on multiple domains including legal texts, already surpasses BM25 performance even without fine-tuning.

After fine-tuning, the Bi-Encoder shows significant improvement, reaching approximately \textbf{97\% Exist@90} and \textbf{61\% MRR@10}. Based on these results, we select the version fine-tuned for 11 epochs as our final retrieval model.

\subsection{Re-ranking Stage with Cross-Encoder and Negative Sampling Strategies}

The re-ranking stage is critical to the overall performance of the retrieval pipeline, as it determines the final ranking of candidate documents. We use PhoRanker as the Cross-Encoder model due to its pre-training on Vietnamese data and relatively fast inference speed, which make it suitable for our task. However, Cross-Encoders do not include built-in loss functions tailored for question-answer formats, which can introduce bias during training if negative samples are not properly selected. To address this, we incorporate a negative example mining strategy.

We leverage the fine-tuned Bi-Encoder from the retrieval stage to generate negative samples. Specifically, for each question in \texttt{train\_df}, the Bi-Encoder retrieves the top 90 candidate documents. Based on these candidates, we explore three types of negative sampling strategies:

\begin{itemize}
    \item \textbf{Hard Negative Mining}: The correct answers are removed from the candidate list, and the top-$n$ highest-ranked remaining candidates are selected as negatives.
    \item \textbf{Semi-Hard Negative Mining}: After removing the correct answers, $n$ candidates are randomly selected from the remaining ones.
    \item \textbf{Easy Negative Mining}: $n$ examples are randomly sampled from the entire corpus (excluding correct answers), without reference to retrieval scores.
\end{itemize}

We experiment with $n \in \{2, 5, 10\}$ for each mining strategy. For strategies involving randomness, we test across three random seeds: 28, 42, and 2025. All Cross-Encoder models are fine-tuned for 2 epochs using a learning rate of $2 \times 10^{-5}$.

\begin{table}[h!]
\centering
\caption{Table comparing the performance of the reranker model when fine-tuned on different types of negative examples}
  \label{tab:comparison2}
  \begin{tabularx}{\textwidth}{Xc}
    \toprule
    \textbf{Model}  & \textbf{MRR@10} \\
    \midrule
    PhoRanker & 0.5584 \\
    \midrule
    Hard negative mining, $n=2$ & 0.2689 \\ 
    Semi-hard negative mining, $n=2$, seed = 28 & \textbf{0.7681} \\ 
    Semi-hard negative mining, $n=2$, seed = 42 & \textbf{0.7666}\\ 
    Semi-hard negative mining, $n=2$, seed = 2025 & \textbf{0.7699} \\ 
    Easy negative mining, $n=2$, seed = 28 & 0.5481\\ 
    Easy negative mining, $n=2$, seed = 42 & 0.5637\\ 
    Easy negative mining, $n=2$, seed = 2025 & 0.5375 \\ 
    \midrule
    Hard negative mining, $n=5$ & 0.4796 \\ 
    Semi-hard negative mining, $n=5$, seed = 28 & \textbf{0.7821} \\ 
    Semi-hard negative mining, $n=5$, seed = 42 & \textbf{0.7791}\\ 
    Semi-hard negative mining, $n=5$, seed = 2025 & \textbf{0.7809} \\ 
    Easy negative mining, $n=5$, seed = 28 & 0.5701  \\  
    Easy negative mining, $n=5$, seed = 42 & 0.5675 \\ 
    Easy negative mining, $n=5$, seed = 2025 & 0.5703  \\ 
    \midrule
    Hard negative mining, $n=10$ & 0.6751 \\ 
    Semi-hard negative mining, $n=10$, seed = 28 & \textbf{0.7911}\\ 
    Semi-hard negative mining, $n=10$, seed = 42 & \textbf{0.7892}\\ 
    Semi-hard negative mining, $n=10$, seed = 2025 & \textbf{0.7900} \\ 
    Easy negative mining, $n=10$, seed = 28 & 0.5940 \\  
    Easy negative mining, $n=10$, seed = 42 & 0.6074 \\ 
    Easy negative mining, $n=10$, seed = 2025  & 0.5790 \\ 
    \bottomrule
  \end{tabularx}
  \label{tab5}
\end{table}

As shown in Table~\ref{tab:comparison2}, four key insights emerge: random seed variations yield consistent performance outcomes, semi-hard negative mining reliably achieves the best results, all approaches benefit from increasing the number of negative samples, and hard negative mining surpasses easy negative mining only when \(n=10\).

\textbf{Is the sole benefit of mining negative samples just to avoid bias?} If the only purpose were to mitigate bias and simply increase the cosine similarity scores of positive samples, then employing easy negative mining would be sufficient. However, as indicated in Table \ref{tab5}, easy negative mining never performs as well as semi-hard negative mining. Mining negative samples from the candidate set not only encourages higher cosine similarity scores for positive samples but also drives down the scores for potential candidates that are not the correct answers, thus accelerating model convergence.

\textbf{Why does hard negative mining perform poorly?} We suspect that hard negatives are often too similar to true positives, making them difficult for the model to distinguish and potentially introducing noise due to overlapping features. When training the Cross-Encoder with BCEWithLogitsLoss using binary labels (0 and 1), very challenging negatives with high initial similarity scores (e.g., around 0.9) can lead to large loss values. This, in turn, causes strong gradients that may abruptly shift the model’s weights, resulting in unstable training and slower convergence.

Using PhoRanker, we measured the initial cosine similarity between each negative example and its corresponding question under three sampling strategies: hard, semi-hard, and easy negatives. This analysis was conducted with $n = 2$ negative samples per question. Table~\ref{tab6} summarizes the distribution of these samples across four similarity intervals: less than 0.5, between 0.5 and 0.8, between 0.8 and 0.9, and greater than or equal to 0.9. The final column reports the average similarity score for each strategy.

\begin{table}[h!]
  \centering
  \caption{Distribution of negative samples by cosine similarity with the question}
  \label{tab6}
  \renewcommand{\arraystretch}{1.1}
  \begin{tabular}{
      >{\columncolor{gray!12}}l
      >{\columncolor{white}} S[table-format=1.4]
      >{\columncolor{gray!12}} S[table-format=1.4]
      >{\columncolor{white}} S[table-format=1.4]
      >{\columncolor{gray!12}} S[table-format=1.4]
      >{\columncolor{white}} S[table-format=1.4]
    }
    \toprule
    \textbf{Type} &
      {$<0.5$} &
      {$0.5$--$0.8$} &
      {$0.8$--$0.9$} &
      {$\ge 0.9$} &
      {Mean} \\
    \midrule
    Hard      & 0.2833 & 0.1021 & 0.1059 & 0.5087 & 0.6806 \\
    Semi-hard & 0.7944 & 0.0718 & 0.0474 & 0.0864 & 0.2072 \\
    Easy      & 0.9996 & 0.0002 & 0.0001 & 0.0001 & 0.0008 \\
    \bottomrule
  \end{tabular}
\end{table}

As shown in Table~\ref{tab6}, over 60\% of the hard negative samples have a cosine similarity score of 0.8 or higher, indicating that these examples are highly similar to the true positives. Such high similarity can confuse the model, especially when using binary labels with \texttt{BCEWithLogitsLoss}. The loss function responds with large gradients when trying to push these near-positive samples toward a zero score. These abrupt updates can lead to unstable training, slower convergence, or poor generalization. In contrast, semi-hard and easy negatives are more distinguishable, allowing the model to learn more reliably. This distributional insight helps explain the observed drop in performance when training with hard negatives.

\textbf{Why does hard negative mining perform well when n=10?}
Based on the described hard negative sampling strategy, increasing $n$ leads the model to include relatively easier negatives (i.e., semi-hard negatives). In the extreme case where $n=90$, hard negative mining effectively becomes semi-hard negative mining. This balance explains why the performance of hard negative mining improves with a larger $n$.

\section{Conclusion}
This study introduces a two-stage framework for legal document retrieval, addressing LLM limitations in specialized domains. By integrating a fine-tuned Bi-Encoder (97\% Exist@90) for efficient candidate retrieval and a Cross-Encoder re-ranker (79.11\% MRR@10) optimized via semi-hard negative mining, our method outperforms traditional approaches like BM25. Despite its simplicity, our model achieves same performance compared to competitors that rely on complex ensemble methods. The streamlined pipeline highlights the importance of domain-specific adaptation, balanced negative sampling, and hybrid architectures for mitigating bias and enhancing efficiency. Validated on a Vietnamese legal corpus, this work provides a scalable blueprint for robust retrieval-augmented systems in complex, knowledge-intensive domains.

\section*{Acknowledgement}\label{conclusion}
This research is funded by Vietnam National University HoChiMinh City (VNU-HCM) under the grant number DS2024-26-01.

\sloppy
\printbibliography

\end{document}